\documentstyle[epsf, prd, aps, floats]{revtex}
\newcommand{\be}{\begin{equation}}
\newcommand{\ee}{\end{equation}}
\newcommand{\bea}{\begin{eqnarray}}
\newcommand{\eea}{\end{eqnarray}}

\begin{document}
\twocolumn[\hsize\textwidth\columnwidth\hsize\csname@twocolumnfalse\endcsname
\author{Stephon Alexander $^{1,2}$, Louis Crane $^{2}$, Marni Sheppeard $^{2,3}$}
\date{\today}
\address{1) Dept of Physics and SLAC,
Stanford University, 2575 Sand Hill Rd, Menlo Pk, CA, U.S.A.}
\address{2) Perimeter Institute, Waterloo, ON, Canada}
\address{3) Dept of Physics and Astronomy, University of
Canterbury, Christchurch, New Zealand}
\title{The Geometrization of Matter Proposal in the Barrett-Crane Model
and Resolution of Cosmological Problems}
\maketitle
%--------------------------------------------------------------------------
\begin{abstract}
We give an overview of the current issues in early universe
cosmology and consider the potential resolution of these issues in
an as yet nascent spin foam cosmology. The model is the
Barrett-Crane model for quantum gravity along with a
generalization of manifold complexes to complexes including
conical singularities.
\end{abstract}
]
\renewcommand{\thefootnote}{\arabic{footnote}}
\setcounter{footnote}{0}
%--------------------------------------------------------------------------
\bigskip
\section{Introduction} \bigskip

The Barrett-Crane model \cite{BC} \cite{Smolin1995} is a
constrained topological state sum model for quantum gravity.
Recently \cite{Matter} it was proposed that this model might
incorporate matter and gauge interactions if the condition on
triangulations to be manifolds were relaxed. That is, conical
singularities would act as seeds of matter in the quantum geometry
of the state sum. The purpose of this paper is to examine the
consequences of this proposal, and of spin foam models in general,
for early universe phenomenology. We feel that although the BC
model, and the conical matter proposal in particular, are not yet
sufficiently well understood, the connections are provocative.
Conversely, one would like to use cosmology as a guiding principle
for developing experimental techniques in spin foam models that
could be used to explain and predict real observations.

The point of this paper is that natural approximations motivated
by the conical matter proposal (CMP) seem to shed light on the
{\em full} range of puzzles of the early universe. Although each
argument is in need of further substantiation, the overall picture
seems striking in itself.

To clarify this point, and since the communities of researchers in
spin foam models and early universe phenomenology are not
generally aware of each other's work, this paper begins with a
self contained introduction to the range of early universe
phenomenological issues. We then give a nontechnical introduction
to the categorical state sum program and the BC model in
particular, together with the conical matter proposal. We conclude
with a discussion of a range of phenomenological problems which
have plausible solutions in the state sum picture.

Although we have tried to make the paper accessible to
cosmologists by giving physical explanations for the state sum
models as far as possible, they do make use of branches of
mathematics not generally familiar to physicists, so interested
readers are strongly recommended to consult the references.

We believe that the current form of the model is likely to be less
important than the fact that it leads to calculations that probe
poorly understood phenomena.

%--------------------------------------------------------------------------
\bigskip
\section{Current Issues in Early Universe Cosmology}
\bigskip

\subsection{The Flatness Problem}
The Standard Big Bang predicts that in an expanding space-time
$\Omega=1$ is an unstable fixed point. However, present
observations confirm that our universe is flat, i.e. $\Omega=1$ to
within 10 percent. This means that $\Omega$ had to be very close
to one in the past. For example, at the era of nucleosynthesis, we
are constrained to have $|\Omega-1| < O(10^{-16})$. This is an
extreme fine tuning of initial conditions. Unless initial
conditions are chosen very accurately, the universe soon
collapses, or expands quickly before structure can be formed. The
suggestion discussed below of a topological phase may shed some
light on the flatness problem.

\subsection{The Horizon Problem}
If we look back at the surface of last scattering we see
homogeneity in the cosmic microwave radiation across distances
subtending $10^{28} cm$. However the size of correlations
described by causality at the surface of last scattering is
predicted by SBB to be $d_{cor} \sim 10^{23}cm$, a large
discrepancy between observations and theory.  Therefore, the SBB
provides no causal way to produce correlations to establish
homogeneity on such large scales that are observed in the cosmic
microwave background (CMB). Inflation was designed partially to
solve this problem. We suggest a solution to the horizon problem
where the initial conditions of the universe may already have
established correlations that were homogeneous.

\subsection{Baryon Asymmetry}
Current observations tell us that most of the universe is made up
of non-baryonic degrees of freedom such as dark energy (2/3), dark
matter (1/3) and only 5\% baryonic matter.  An even more striking
observation of the smallness of baryon density is given by the
ratio of baryons to photons in the CMB. Big Bang nucleosynthesis
as well as anisotropies in the CMB yield the ratio
\begin{equation} \frac{n_{b}}{n_{\gamma}} = 6.1 \times 10^{-10}
\end{equation}

Why is this number so small, yet nonzero? If equal numbers of
particles and antiparticles had been created in the early
universe, they would have annihilated in pairs. It is well known
that the Standard Model at the level of renormalizable terms
possesses no interactions that can violate either baron number or
lepton number violation. In 1967 Sakharov \cite{Sakh} studied how
the baryon asymmetry could arise. He summarized the three
conditions required of any process that would lead to baryon
asymmetry:
\begin{itemize}
\item Baryon number violation must occur in the fundamental
interactions
\item CP violation
\item Local violation of thermal equilibrium, including an Arrow of
Time.
\end{itemize}
We will observe that the conical matter proposal plausibly
satisfies the Sakharov conditions.

\subsection{Dark Matter}
Our current understanding of structure formation necessitates the
existence of pressureless non-baryonic dark matter.  Observations
on galactic and cosmological scales reveal that 22 percent of the
matter in our universe consists of non-baryonic dark matter
(NBDM). NBDM does not interact with radiation so it can not be
detected by standard astronomical means.  More importantly, there
are a few candidates for NBDM, such as SUSY neutralinos, axions,
shadow universes and Planck mass black hole remnants.  These are
plausible candidates because they are fields which are weakly
interacting.  However, they are all based on extensions of the
Standard Model and will likewise suffer the same type of initial
condition fine tuning as scalar fields coupled to gravity in a
cosmological setting. Below, we suggest a plausible new candidate
for DM in our model, namely higher genus conical matter, and also
a relationship between these and Primordial Black Holes (PBH).

A very concrete challenge to theoretical physics is posed by the
recent detailed observations of dwarf galaxies \cite{Dwarf}. The
density of dark matter in dwarf galaxies as a function of radius
can be computed very accurately because they are dominated by dark
matter even near their cores. The observed distribution does not
agree with the theoretical prediction for non-interacting dark
matter in that it lacks a central cusp.

\subsection{Inflation}
The idea that the early universe underwent a phase of exponential
growth has many attractions, such as its ability to reproduce the
acoustic peaks in deviations to the CMB power spectrum.
Unfortunately, current theoretical models require an unmotivated
fine tuning of the potential of an as yet hypothetical scalar
field. We will propose that the unusual thermodynamics of conical
matter provides a plausible mechanism for inflation, and the
decoupling of the higher genus particles could provide an exit
scenario.

%--------------------------------------------------------------------------
\bigskip
\section{The Standard Big Bang} \bigskip

The Standard Big Bang Scenario is based on three observational
pillars: primordial nucleosynthesis, isotropy and homogeneity seen
in the cosmic microwave background and the Hubble redshift
relation.  SBB is the simplest general relativistic scenario which
predicts these observations. Despite its success the SBB suffers
from other observational and theoretical problems.  By now these
problems are well known \cite{RB}. A key issue for model builders
is to simultaneously resolve these problems while keeping the SBB
in the regime where it predicts observations consistently.  The
main theoretical underpinning of SBB is the use of Einstein's
general theory of relativity minimally coupled to a gas of
particles (a hot perfect fluid in thermal equilibrium).  However,
at high curvatures and hence early times, this approximation is no
longer valid and the SBB needs to be modified.  Some of the main
problems of the SBB were partially solved by the inflationary
paradigm. Without changing general relativity but by relaxing the
assumption of the equation of state of matter to include a
contribution from a quantum field, the inflaton, the inflationary
paradigm was able to simultaneously resolve the problems of the
SBB and even provide a causal mechanism of generating a scale
invariant power spectrum. The conventional wisdom stemming from
QFT is to minimally couple gravity to quantum fields, which
realizes inflation, although it has been shown that this
assumption  breaks down in the early universe. Unfortunately,
inflation does not shed much light on the dark energy and dark
matter problems.  In fact, inflation seems to be pointing to the
roots of its own demise, namely the transplanckian problem.
Briefly, the transplanckian problem of inflation is that
structures on scales of cosmological interest today were
generically generated deep within the Planckian regime where the
assumption of a scalar field minimally coupled to gravity breaks
down.

With these issues in mind, we consider a different approach. There
are features, to be discussed below, of the categorical state sum
(CSS) approach to quantum general relativity which are useful for
proposing new solutions to some of the cosmological problems,
which interestingly are not soluble by inflation or other
modifications to the SBB.

The problems we will discuss are inflation, dark matter and the
flatness, horizon and chirality problems, as these appear to have
distinctive realizations in the BC model of quantum gravity and
yield a new perspective as to how quantum gravity can resolve
these problems in cosmology without resorting to the logic of fine
tuning an effective field theory.

%--------------------------------------------------------------------------
\bigskip
\section{State Sum Models, Quantum Gravity and Conical Matter}
\bigskip

The basic idea of the state sum approach to quantum gravity
\cite{reviewP} is that quantum geometry is a superposition of
discrete quantum processes of the same form as Feynman diagrams.
Geometries are quantized by using representation theory to obtain
Hilbert spaces on which geometric quantities act as operators.
Thus, a categorical state sum is a discretized version of a
Feynman vacuum, where the fields and vertices correspond to
Lorentzian geometry.

The diagrams are not considered to be embedded in a background
classical spacetime. Rather, the combinatorial structure of the
diagrams themselves yields spacetime, and the quantum fields on it
represent a sum over metrics. The structure of spacetime in this
picture is given by a simplicial complex, in which the individual
Feynman diagrams are quantum geometries of four dimensional
simplices in Minkowski space.

In other words, quantum geometry in this approach is represented
by families of Hilbert spaces on which the sort of quantities
typically measured in classical geometries act as operators. The
most basic geometric quantities are bivectors, i.e. skew symmetric
rank two tensors, which describe oriented area elements. Utilizing
their expedient quantization, we define the other geometric
quantities in terms of these bivectors, which are attached to the
faces of a triangulation.

In the quantization procedure of the BC model \cite{BC}, bivectors
are represented by unitary representations of the Lorentz algebra.
The bivectors on faces and tetrahedra are constrained to be
simple, i.e. to correspond to oriented area elements. This has a
natural quantization in the restriction to the balanced unitary
representations. We construct the model by using harmonic analysis
to describe these representations.

Specifically, the balanced representations are given by families
of functions on the hyperboloid $H^3$. The projection onto the
balanced representation with real parameter $\rho$ is given by
\begin{equation}
\frac{1}{2 \pi^{2}} \int_{H^3} K_{\rho}(x,y) h(y) d y
\end{equation}
where
\begin{equation} K_{\rho}(x,y) = \frac{\sin \rho r(x,y)}{\rho \sinh r(x,y)}
\end{equation}
for $r(x,y)$ the hyperbolic distance. The $10j$ symbol which
corresponds to the quantum geometry of a Lorentzian $4$-simplex
$s$ is represented by the integral
\begin{eqnarray}
(10j)_{s} = \int d x_{1} d x_{2} d x_{3} d x_{4} d x_{5}
K_{\rho_{1}} (x_{1},x_{5}) \\ \nonumber K_{\rho_{2}} (x_{1},x_{4})
K_{\rho_{3}} (x_{1},x_{3}) K_{\rho_{4}} (x_{1},x_{2})
\\ \nonumber K_{\rho_{5}} (x_{2},x_{5}) K_{\rho_{6}} (x_{2},x_{4})
K_{\rho_{7}} (x_{2},x_{3}) \\ \nonumber K_{\rho_{8}} (x_{3},x_{5})
K_{\rho_{9}} (x_{3},x_{4}) K_{\rho_{10}} (x_{4},x_{5})
\end{eqnarray}
depending on the $10$ face labels. The complete state sum as a sum
(and integral) over labellings $c$ of the triangulation takes the
form
\begin{equation}
{Z}_{BC} = N \sum_{c} \prod_{t} A_{t} \prod_{f} {\rho_{f}}^{2}
\prod_{s} (10j)_{s}
\end{equation}
where for the tetrahedron $t$
\begin{equation}
A_{t} = \int d x K_{\rho_{1}} (x) K_{\rho_{2}} (x) K_{\rho_{3}}
(x) K_{\rho_{4}} (x)
\end{equation}
See \cite{BC} and \cite{Perez} for details. Since we think of the
individual triangulations themselves as Feynman diagrams for a
more fundamental theory called Group Field Theory \cite{GFT}, we
want to make a summation over triangulated complexes to produce
the full theory.

The actual construction of the theory depends on the idea that
general relativity can be described as a constrained version of BF
theory \cite{reviewP} \cite{Smolin1995}, which has the action
\begin{equation} S = \int B \wedge F
\end{equation}
where $B$ is a $2$-form and $F$ the curvature of a connection $A$.
In fact, the state sum of the BC model is a constrained version of
the topological CKY model of \cite{CKY}, which is a quantization
of BF theory. The quantization in \cite{CKY} also has bivectors as
variables. The constraint which transforms the CKY model to the BC
one is the restriction to bivectors that are simple, i.e. that
correspond to oriented area elements rather than superpositions of
them.

The steepest descent condition (the analog of the classical
equation of motion) for the CKY model corresponds to the condition
of flat geometry, while the constraint introduced in the BC model
converts this to an analog of Ricci flatness, i.e. to a
discretization of Einstein's equation. In order to make the sum
over triangulations finite, it is very tempting to consider the
hypothesis of a 'phase transition' in the early universe, in which
the constraint of the BC model emerges as a result of a kind of
collapse of the wave packet of the unmeasured universe. The form
of the constraint mentioned above is suggestive of that, in that
it suppresses superpositions of simple bivectors. We warn the
reader that there is as yet no model of how this could come about,
and that the work in quantum information theory \cite{Dorit} which
suggests it is also lacking a mechanism. A plausible model for the
transition is outlined below, and can be thought of as an attempt
to make precise the 'topological phase' suggestion of Witten
\cite{Witten}.

At this point, we discover that the class of triangulated spaces
which occur as Feynman graphs for GFT is broader than the class of
triangulated four dimensional manifolds. As explained in
\cite{Matter}, it is possible to have edges and vertices whose
points are conical singularities. The edge points can appear as
cones over closed surfaces, while the vertices can appear as cones
over general three manifolds, which contain the conical
singularities over all surfaces corresponding to edges incident on
the vertex.

Let us restate this. Every Feynman diagram for the GFT approach to
the BC model has the topology of a four manifold containing a web
of conical singularities taking the form of a graph, the points of
whose edges have neighborhoods which are cones over surfaces,
while the vertices of the graph have neighborhoods which are cones
over 3-manifolds with boundaries, the boundary components fitting
to the boundary surfaces on the incident edges. This is a theorem
of combinatorial topology; it just summarizes the ways that the
4-simplices in the model can be glued together.

Differently put, in passing from physical theories described by
differential equations to discrete models, we find that the
natural class of spacetimes to consider has broadened, from smooth
manifolds to what mathematicians would call PL pseudomanifolds.

\begin{figure}
\centering
\epsfbox{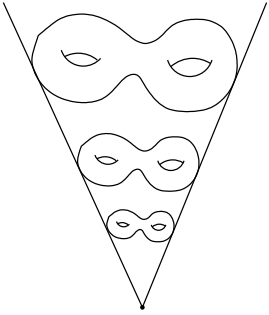}
\caption{cone over surface}
\end{figure}

\subsection{How to Get Matter}

The conical matter proposal (CMP) is to consider a) the conical
singularities on edges as generating particles which propagate
through space, and b) the conical singularities on vertices as
interaction vertices.

Perhaps it is useful to try to picture this. Feynman taught us to
think of the vacuum as full of processes described by Feynman
diagrams. Instead, we are proposing a picture where spacetime is
full of edges, each point of which is a cone over a surface (fig
1), joined at vertices which are cones over three manifolds with
boundary; in the simplest case, cones over link complements (fig
2). These would fit together into complexes with a combinatorial
structure analogous to Feynman diagrams (fig 3).

\begin{figure}
\centering
\epsfbox{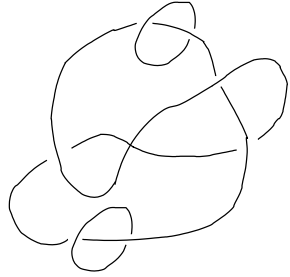} \caption{three manifold with
boundary components}
\end{figure}

It is interesting to note that this is a direct extension of the
most popular technique for adding matter in 2+1 dimensional
gravity to the 3+1 setting. In 2+1 dimensions, matter is added in
the form of conical curvature singularities. In the quantum
theory, this is expressed by adding punctures to the Riemann
surfaces of the spatial slices. In 2+1 dimensions, a conical
singularity is not a topological defect; that is a new feature in
3+1 dimensions.

\begin{figure}
\centering \epsfbox{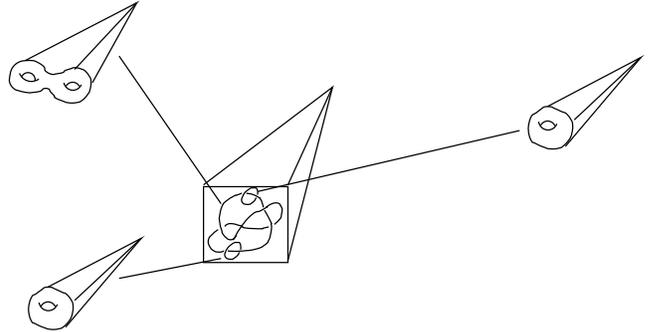} \caption{conical vertex as Feynman
vertex}
\end{figure}

In comparison to other fundamental physical theories involving
matter, the CMP has one advantage: matter is naturally included in
the theory of quantum gravity, rather than added by hand. There is
{\em no} new element; neither a gauge group, nor extra dimensions,
nor a topology on a compact manifold. The surfaces are not
insertions into the space-time; they are only descriptions of part
of its topological structure. It is therefore highly remarkable,
as we will explain, that the most natural approximation scheme
available suggests that the Standard Model may emerge from it.

In order to investigate the implications of the CMP for low energy
physics, two approximation techniques had to be adopted. It was
assumed that the low energy behavior, both at conical
singularities along edges and at vertices, would be dominated by
the flat metrics on the regions around the singularities. It was
further assumed that the state propagating along an edge would
correspond to the states of the topological quantum field theory
of the Lorentz algebra on the boundary surface of the conical
singularity.

We should justify these assumptions by saying that any
configuration around the singularity which required curvature
would therefore be very massive, and that since the CSW TQFT
\cite{Witten} has flatness for an equation of motion, it is a
quantization of the flat geometries. We regard this approach as
preliminary, and hope that further progress in understanding the
CSS models will provide a better underpinning for it.

Another way of thinking of this approximation is as a novel form
of the holographic principle \cite{Fotini} \cite{Smolin}. The
conical regions are analogous to black holes with more complex
topology. The spaces of states on them is described by the states
of CSW on their boundaries.

The implications of these assumptions are interesting. The
classification of flat Lorentzian  metrics over a cone is
equivalent to the classification of hyperbolic metrics on its
boundary. This means that the space of flat geometries on the cone
over a surface is described by its Teichmuller space. If we take
this as confirmation that the effective low energy states of the
theory are given by the CSW theory for the q-deformed Lorentz
algebra on the once punctured torus, then this has the structure
of the quantum group $U_q(su(2))$, which as an algebra has the
structure
\begin{equation} U_q(su(2)) = \bigoplus_{n = 1}^{\infty} Mat_{n}(\mathbf{C})
\end{equation}
of a direct sum of one matrix algebra of each dimension. The
unitary part of this reproduces the gauge group of the Standard
Model in a manner analogous to the Connes-Lott model
\cite{Connes}, if truncated after the first three terms. Since
noncommutative geometries arise naturally in the description of
singular spaces, in particular of spaces which admit descriptions
as quotients of more regular spaces, as conical singularities do,
it would be reasonable to try to connect our picture more directly
with the noncommutative geometry approach.

The truncation can either be done by picking a third root of unity
for q as in \cite{Coq}, or one can simply consider the possibility
that particles with $su(4)$ or higher quantum numbers are
extremely massive, a possibility which is at least open to study.

Consider the easiest nontrivial cobordism ($3$-manifold with
boundary) linking three tori. The most obvious interaction between
them corresponds to multiplication in the algebra $U_q(su(2))$, so
that the states on the torus get a natural interpretation as gauge
bosons for the Standard Model. The question of the orientation of
the surface needs careful study. Thus, we have candidates for
photons, gluons etc. in our model, with the right sort of
interactions. In order to determine the masses, propagation and
spins of these, we will need better approximations.

The problem of incorporating fermions is more difficult to
understand. The space of states on the cone over a Klein bottle
seems a natural candidate, but more developed approximations will
be necessary to study this question. The one thing we can say is
that one might look for chiral asymmetry in the interaction
vertices represented by $3$-manifolds with boundary; in
particular, link complements. Given the discovery of neutrino
masses, this at least seems a plausible approach.

The case of flat metrics over a vertex is more complex. Hyperbolic
three manifolds with complete metrics correspond to Kleinian
groups and play an important role in the topology of three
dimensional manifolds.

A number of important facts are known about hyperbolic
$3$-manifolds with boundary. For instance, if any of the boundary
components have genus greater than 1, then the complete hyperbolic
metric has infinite volume, which in fact grows exponentially as
we approach the end, while the genus 1 (tori and Klein bottle)
have finite total volume. Another rather obvious fact is that
$3$-manifolds with boundary give natural examples of topological
objects which are not equivalent to their mirror images, such as
the class of $3$-manifolds arising from taking the complement of
an ordinary link in $S^3$.

A simple physical interpretation of these facts is that at low
energy the states corresponding to singularities over higher genus
surfaces decouple both from the genus 1 states and from one
another. The states on tori and Klein bottles remain interacting,
giving rise to a world of gauge bosons and fermions, as
conjectured above. The higher genus states, therefore, were
interacting with the genus 1 states in the early universe, but
decoupled at some later stage.

The assumptions that we can take the TQFT Hilbert space on the
boundary of a conic singularity as the physical Hilbert space of
its states, and the CSW amplitude for the three dimensional
boundary of a conical vertex as its amplitude, have the further
interesting implication that the dimension of the Hilbert space
goes up exponentially with the area of the boundary, while the
amplitude of the vertex goes up exponentially with its volume.

The first observation is closely analogous to the observation that
the dimension of the Hilbert space on a punctured sphere goes up
exponentially with the number of punctures, which led Smolin
\cite{Smolin1995} to conjecture a connection between TQFT and
quantum gravity. Higher genus surfaces are very similar to
surfaces with many punctures in TQFT. The second observation has
interesting implications for a universe with higher genus conical
dark matter.

%--------------------------------------------------------------------------
\bigskip
\section{Connections} \bigskip

The picture which thus arises from considering CSS models,
including the suggestion of a TQFT-BC type phase transition and
the CMP, makes contact with a range of puzzles in early universe
phenomenology.

Let us summarize the picture of the history of the universe which
our model seems to suggest (fig 4). There would be an early (or
rather sub-Planckian) phase, in which the universe would be
modelled by a topological quantum field theory, but with conic
singularities included in the manifold. This phase would be a
substitute for the initial singularity of the SBB model. It would
be followed by a phase of quantum gravity, in which genus 1 and
higher genus conic singularities would be interacting, while the
universe expanded and cooled. Next would come a decoupling, in
which further interactions involving higher genus singularities
would be suppressed by topological obstructions, leaving an
interacting world composed of genus 1 singularities.

\begin{figure}
\centering
\epsfbox{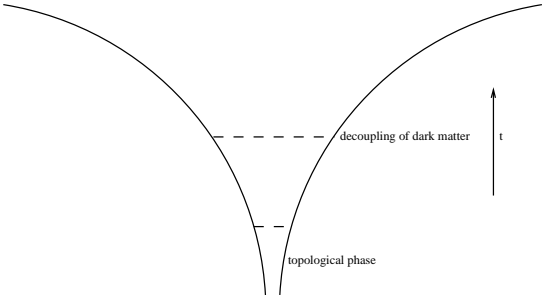}
\caption{history of early universe}
\end{figure}

We now discuss how features of this CSS program relate to the
specific phenomenological issues outlined above.

\subsection{The Cosmological Constant and Quantum Groups}

Since it is very natural to pass from classical to quantum groups
in constructing categorical state sums, and in particular since
the q-BC model is well behaved \cite{qBC}, our discussion above
easily accommodates a cosmological constant, and indeed may even
require it. The representation theory of the quantum Lorentz
algebra seems to give a quantum geometry of space with constant
curvature very similar to the quantum geometry of flat space
provided by the unitary representations of the usual Lorentz
algebra. It also provides additional regularization, which may be
useful for suitable normalizations; in particular, for the
contribution of higher genus singularities.

\subsection{The Flatness and Horizon Problems and the Topological Phase}

The idea of a 'phase transition' from the BF TQFT to a BC type
model suggests an approach to the flatness problem, since the
equations of motion for the BF theory are exactly flatness. Also,
since a TQFT has no light cones, the distribution of conical
matter would be random, transforming into a thermal distribution
when the metric materialized. This also suggests an approach to
the horizon problem: we regard the substitution of a topological
phase for a point-like singularity as a positive feature of our
model.

The phase transition might arise as a result of coarse graining of
the topological universe, which at the origin of time fluctuated
into a combination of quantum variables corresponding to a
sufficiently large 'size'. Small black holes would cause a loss of
phase information, which would mediate a transformation to a
regime described by a sum over simple bivectors only, which, as we
explained above, has GR as a classical limit. The GR equations of
motion would then prevent the escape of information from the black
holes. We believe this process could be modeled using techniques
similar to those of \cite{Dorit}.

Although this proposal is self-consistent, it has a disturbing
chicken versus egg quality. The question of the phase transition
is the point which most strongly suggests to us that still deeper
theoretical constructions will ultimately be needed in this
approach.

There is at least one toy candidate for investigating the phase
transition within the timelike q-BC model. The constraint
condition in terms of representations of the (quantum) Lorentz
group is that a real spin label $k \in \frac{1}{2} \mathbf{Z}$
should equal zero. One could consider a state sum cut off at any
maximum $k$, the topological theory being recovered when $k
\rightarrow \infty$. If a spacelike model is considered instead,
this 'thermal' parameter becomes continuous. This idea is
strengthened by \cite{Capo} and \cite{Livine} in which a
generalized action of the form
\begin{equation}
S(B,A,\phi ,\mu) = \int B^{IJ} \wedge F_{IJ} - \frac{1}{2}
\phi_{IJKL} B^{IJ} \wedge B^{KL} + \mu H
\end{equation}
is discussed, where $A$ is an $SO(3,1)$ connection, $\mu$ and
$\phi$ Lagrange multipliers such that $\phi$ satisfies
$\phi_{IJKL} = - \phi_{JIKL} = - \phi_{IJLK} = \phi_{KILJ}$ and
\[ H = a_{1} \phi_{IJ}^{IJ} + a_{2} \phi_{IJKL} \epsilon^{IJKL}
\]
The Immirzi parameter $\gamma$ is introduced by
\begin{equation} \frac{a_{2}}{a_{1}} = \frac{1}{4} ( \gamma -
\frac{1}{\gamma} )
\end{equation}

In the Lorentzian case the generalized action always corresponds
to the $q$-deformed Barrett-Crane model. When $\gamma = 0$ the
topological theory is recovered.

\subsection{Inflation}

Let us now briefly recall our current understanding of the
Bekenstein bound for the entropy of a black hole in terms of
TQFTs. The hyperbolic area of a uniform Riemann surface,
normalized to a constant curvature of $-1$, depends only its
genus. Such a surface may be cut up into cylindrical and trinion
pieces, as shown in figure 5. These pieces are clearly
homeomorphic to the punctured spheres that represent black hole
horizons. In a TQFT that assigns a Hilbert space to each such
puncture, the invariant $Z$ thus depends on the hyperbolic area
and increases exponentially with the number of punctures. A simple
corollary of this is that the entropy of the black hole scales
with horizon area. For an overview of entropy bounds see
\cite{Smolin} and references therein.

\begin{figure}
\begin{center}
\epsfbox{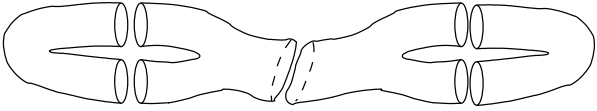}
\end{center}
\caption{surface decomposition in $2D$ TFT}
\end{figure}

In an analogous simple analysis of the thermodynamics near
vertices in the conical matter proposal, one considers the
$3$-dimensional topological CSW invariant on the hyperbolic link
complement. Now it so happens that there is a strongly supported
mathematical conjecture due to Kashaev \cite{Kashaev}
\cite{Murmur} which states that the closely related colored Jones
invariant $J_{N}$ is asymptotically related to the normalized
hyperbolic volume $V(L)$ of a link complement in $S^{3}$ by
\begin{equation}
2 \pi \log |J_{N}(L)| \sim N V(L) \hspace{2cm} N \rightarrow
\infty
\end{equation}
where $q = e^{2 \pi i / N}$ and $N \in \mathbf{Z}$.

Assuming that it is reasonable to define thermodynamic quantities
using our partition function, this conjecture seems to suggest
that the contribution of the conical singularities to the pressure
(normalized with respect to energy density) is given by
\begin{equation}
\frac{P}{\rho} = - \frac{2 \pi}{N} \frac{\partial log J}{\partial
V} \sim -1
\end{equation}

A $P / \rho \leq -1/3$ drives inflation, so this result is in
accord with an inflation scenario. In this new scenario {\em no}
inflaton field is introduced. Inflation arises naturally as a
result of the negative pressure. Decoupling of higher genus modes
at low temperature could end inflation. Observe that temperature
scales as $N$.

Restating this argument more physically: the free energy of the
vertex, as given by the partition function for the TQFT, is equal
to the Jones polynomial. Since this equals the exponential of the
volume, the system has a negative pressure equal to the energy
density. This is analogous to the situation for conventional
inflation, where the stress-energy of the inflaton field must be
proportional to the Lorentz metric $g_{\mu \nu} =
diag(1,-1,-1,-1)$ by Lorentz invariance.

We are thus led to conjecture that the appropriate configuration
for a vertex inserted into a nearly flat region (yet to be
analysed) will not depend on the choice of a preferred frame. That
is, it will be manifestly Lorentz invariant. If true, this would
give an interesting physical interpretation to Kashaev's deep
mathematical conjecture.

The main assumption which goes into the above argument is that the
behavior of the state sum in a region around a singular vertex can
be well approximated by GR coupled to an effective stress-energy
with pressure as above. Since the state sum reproduces Einstein's
equation as a classical limit, this is plausible, but needs to be
more carefully studied. We have also assumed that the volume of a
spatial slice is proportional to the boundary volume of a
neighborhood of a conical vertex.

A definite weakness of this stage of development of the model is
that we cannot really justify the above assignation of an energy
to conical matter configurations. To investigate this question
further, one might find the 'best' metrics on the union of a
conical singularity and a patch of surrounding nearly flat smooth
spacetime, and compute the ADM energy. At this point we can only
claim that we are making the most natural guesses in arriving at
phenomenologically interesting conclusions.

\subsection{Dark Matter and Higher Genus Singularities}

One potentially exciting use of the model is in describing dark
matter. Currently there are a few dark matter candidates, most of
which are modelled by weakly interacting scalar fields, like the
axion or the neutralino. But these CDM candidates are problematic
for explaining the lack of cuspiness of dwarf galaxies
\cite{Dwarf}. Naively, one could imagine that the weak residual
interactions of higher genus conical matter would cause the halo
near the center of a galaxy to thermalize, thus eliminating the
(empirically nonexistent) cusps. This should be susceptible to
computational study, and the detail with which the halos can be
observed provides a demanding test.

\subsection{Baryon Asymmetry}

The most difficult of Sakharov's conditions is that concerning
baryon and lepton number violating processes. In the early
universe our model would allow baryons or leptons to couple into
higher genus singularities, which would later freeze out. The
necessary asymmetries could arise from the $3$-manifold topology
of fermionic interactions.

\subsection{Higher Genus Matter and PBH Remnants}

Another interesting idea is to realize dark matter as black hole
remnants \cite{remnant} which survive after black holes evaporate.

Higher genus conical matter could easily emerge as PBH remnants.
Their topology would be an obstruction to complete evaporation.
Currently there is no first principle realization of PBH from
quantum gravity, so it will be a major development to make this
connection more concrete.

\subsection{CP Violation and $3$-Manifolds}

We observed above that couplings which arise from $3$-manifolds
with boundary, and can therefore easily have chiral asymmetry, are
also interesting as a possible mechanism for CP violation.

\subsection{The Phase Transition and Variable Speed of Light Theories}

It would be interesting to see whether or not the TQFT can be
considered a mathematically more elegant version of the variable
speed of light idea, which solves the horizon problem but still
lacks a precise quantum gravitational description \cite{VSL}.
Perhaps a more detailed model of the phase transition would
interpolate between acausal and causal propagation.

%--------------------------------------------------------------------------
\bigskip
\section{Conclusions} \bigskip

In summary, spin foam cosmology appears to satisfy the Sakharov
conditions and has the potential to explain a whole host of major
cosmological problems. The computations which go into this picture
are very rudimentary. At the semiclassical level, the problem of
connecting together flat metrics around vertices and edges into
composites for whole spacetimes needs to be studied much more
carefully. The claim that the considerations of phenomenology
discussed here suggest interesting natural questions for research
in the CSS picture, at least, is well grounded.

It seems very plausible to us that the picture we are outlining
here could play a role analogous to the old quantum mechanics. It
has the flavor of a model constructed out of well understood
mathematical physical tools, which bears a reasonable resemblance
to otherwise puzzling phenomenology. It may seem crazy to
substitute discrete categorical diagrams for a continuum
lagrangian, but it may turn out to be not crazy enough! Judging by
historical experience, only a prolonged dialog with phenomenology
will guide us to a theory which is sufficiently crazy.

It is therefore an attractive feature of this model that it
suggests natural procedures for generating refined approximations.
We could use classical GR techniques to study flat or low
curvature metrics on combinations of conical singularities, on
edges and on vertices, and patches of nearly flat smooth spacetime
about these. Calculations based on such approximations could be
combined into models for dark matter or inflationary cosmologies,
among other possibilities, and compared to empirical data.

We have some speculative ideas about the emergence of a deeper
theory. One esthetic drawback to the model we are proposing is
that we begin with spacetime, and produce matter as a sort of
pinch within it. One could wish, rather, for a theory in which
spacetime and matter played dual roles. A hint that such a model
might be possible is that the genus 1 states form the Hopf algebra
object in the category \cite{CraneYetter2} which is a geometric
realization of the braided group of Majid \cite{Majid}. The
representations of the braided group reproduce the category of
representations of the quantum group; this is suggestive of a
deeper model with matter-spacetime duality.

There is enormous scope for investigating early universe
phenomenology in other categorical state sum models. For example,
it appears worthwhile investigating the variation of the
deformation parameter $q$. The relation of $q$ to both a Planck
scale (as a UV cutoff) and a cosmological scale is very suggestive
of a duality such as the aforementioned. One might also consider a
$4$-dimensional model constructed from CSW $S^{3}$ slices of
varying $q$. The discrete parameter $N \in \mathbf{Z}$ labels a
dimensionless 'time' for the topological phase, and might
heuristically enforce a real arrow of time.

Of course, it is much too soon to say anything stronger about the
suggestion that this line of development could lead to a unified
field theory than that it leads to a computational program which
appears tractable and deserves further study.

\bigskip
\noindent {\bf Acknowledgements: } \\
We wish to thank Lee Smolin, Hendryk Pfeiffer, Laurent Freidel,
Hilary Carteret and others at the Perimeter Institute for helpful
conversations, and are grateful to the Institute for its
hospitality. We thank Robert Brandenberger for making many helpful
suggestions from earlier drafts. The last author is partly
supported by the Marsden Fund of the Royal Society of New Zealand.

\bigskip
\bibliographystyle{unsrt}
\bibliography{eu7}
\end{document}